\documentclass[preprint,aps,12pt,showpacs,nofootinbib,tightenlines]{revtex4}
\usepackage{amsmath}
\usepackage{amssymb}
\usepackage{epsfig}
\usepackage{graphicx}
\begin{document}
\def \epjc{  Eur. Phys. J. C }
\def \jpg{  J. Phys. G }
\def \mpla{ Mod.Phys.Lett. A }
\def \npb{  Nucl. Phys. B }
\def \plb{  Phys. Lett. B }
\def \prd{  Phys. Rev. D }
\def \prl{  Phys. Rev. Lett.  }
\def \pr{   Phys. Rep. }
\def \rmp{  Rev. Mod. Phys. }

\newcommand{\bequ}{\begin{equation}}
\newcommand{\eequ}{\end{equation}}

\newcommand{\beq}{\begin{eqnarray}}
\newcommand{\eeq}{\end{eqnarray}}

\newcommand{\bsll}{B \to X_{s} l^+ l^-}
\newcommand{\ssa}{\sin^2\theta_W}
\newcommand{\cca}{\cos^2\theta_W}
\newcommand{\tab}[1]{Table \ref{#1}}
\newcommand{\fig}[1]{Fig.\ref{#1}}
\newcommand{\real}{{\rm Re}\,}
\newcommand{\im}{{\rm Im}\,}
\newcommand{\non}{\nonumber\\ }
\newcommand{\s}{\hat{s}}
\newcommand{\f}{\frac}
\newcommand{\mh}{ m_H}
\newcommand{\mw}{ M_W}

\newcommand{\sm}{{\rm \scriptscriptstyle SM}}
\newcommand{\new}{{\rm \scriptscriptstyle New}}
\newcommand{\np}{{\rm \scriptscriptstyle NP}}

\title{$B \to X_s l^+ l^-$ decay in a Top quark two-Higgs-doublet model}
\author{Zhenjun Xiao}\email{xiaozhenjun@njnu.edu.cn}
\affiliation{Department of Physics and Institute of Theoretical Physics ,
Nanjing Normal University, Nanjing, Jiangsu 210097, P.R.China}

\author{Linxia L\"u} \email{lulinxia@email.njnu.edu.cn}
\affiliation{Department of Physics and Institute of Theoretical Physics ,
Nanjing Normal University, Nanjing, Jiangsu 210097, P.R.China}
\affiliation{Department of Physics, Nanyang Teacher's College, Nanyang,
Henan 473061, P.R.China}
\date{\today}
\begin{abstract}
We calculate the new physics contributions to the rare semileptonic decay
$B \to X_s l^+ l^-$ $(l=e,\mu)$ induced by the charged-Higgs loop diagrams appeared in the top
quark two-Higgs doublet model (T2HDM).
Within the considered parameter space, we found that (a)
the effective Wilson coefficients $\widetilde{C}_{i}^{eff}(m_b)$ ($i =7\gamma, 9V$ and $10A$)
in the T2HDM are always standard model like;
(b) the new physics contributions to $\widetilde{C}_{7\gamma}^{eff}$ and $\widetilde{C}_{9V}^{eff}$
can be significant in magnitude, but they tend to cancel each other;
and (c) the T2HDM predictions for $Br(B \to X_s l^+ l^-)$ agree well with the measured
value within one standard deviation.
\end{abstract}


\maketitle

\section{Introduction}

Flavor changing neutral current (FCNC) $b \to s$ processes
are forbidden at the tree level in the Standard Model (SM). They
proceed at a low rate via penguin or box diagrams. If additional
diagrams  with non-SM particles contribute to such a decay, their
amplitudes will interfere with the SM amplitudes and thereby
modify the rate as well as other properties. This feature makes
FCNC processes an ideal place to search for new physics.

In the past decade, the data of $B \to X_s \gamma$ decay has served
as one of the most important constraints for various new physics
models beyond the SM. At present, the world average, $Br(B
\rightarrow X_s \gamma )=(3.55 \pm 0.26)\times 10^{-4}$
\cite{hfag06}, agrees very well with the standard model prediction
at next-next-to-leading order (NNLO) \cite{gambino}. The magnitude
of the Wilson coefficient $C_{7\gamma}(\mu_b)$ is therefore strongly
constrained by the precision data of $B \to X_s \gamma$, but its
sign is still to be determined through the measurement of $B \to X_s
l^+ l^-$ decay \cite{gambino05}. In Ref.~\cite{gambino05}, the
authors studied  $B \rightarrow X_s l^+ l^-$ decay and found that
the recent experimental data of $Br(B \to X_s l^+ l^-) $ prefer a
SM-like $C_{7\gamma}(\mu_b)$.

In fact, the semileptonic decays $B \to X_s l^+ l^-$ $(l=e,\mu)$ have been extensively
investigated, for example, in the SM \cite{misiak95,buras52}, the two-Higgs doublet
models (2HDM) \cite{huang} or the supersymmetric models \cite{ali02,bobeth05}.
Our goal in the present work is to calculate the new physics contributions to the branching ratio of
$B \to X_s \gamma$ and $B \to X_s l^+ l^-$ decays induced by the charged Higgs loop
diagrams in the top-quark two-Higgs-doublet model (T2HDM) \cite{das1996,kiers,kiers62},
and compare the theoretical predictions in the T2HDM with currently available data.

The outline of the paper is as follows. In section II, we give a brief review for the top-quark
two-Higgs-doublet model and we calculate the new
penguin diagrams induced by new particles and extract out the new
physics parts of the Wilson coefficients or some basic functions
in the T2HDM. In section III,  we present the numerical results of
the branching  ratios of $B \to X_s l^+ l^-$ decay in the SM and the T2HDM, and make
phenomenological analysis. The conclusions are included in the final section.

\section{Theoretical framework}\label{sec:th}

\subsection{Outline of the top quark two-Higgs-doublet model}

The specific model considered here is the top quark
two-Higgs-doublet model (T2HDM) proposed in Ref.~\cite{das1996} and
studied in Refs.\cite{kiers,kiers62}, which is also a special case
of the 2HDM of type III \cite{hou1992}. In this model, the large
mass of the top quark arises naturally in the extension of the SM
since the top quark is the only fermion receiving its mass from the
vacuum expectation value (VEV) of the second Higgs doublet. All the
other fermions receive their masses from the VEV of the first Higgs
doublet.

Let us now briefly recapitulate some important features of the
model of Ref.\cite{das1996}. Consider the Yukawa Lagrangian of the
form:
\beq
{\cal L }_Y = - {\overline{L}}_L \phi_1 E l_R -
{\overline{Q}}_L \phi_1 F d_R - {\overline{Q}}_L
{\widetilde{\phi}}_1 G {\bf 1}^{(1)} u_R -
{\overline{Q}}_L{\widetilde{\phi}}_2 G {\bf 1}^{(2)} u_R + H.c.
\eeq
where $Q_L$ and $L_L$ are 3-vector of the left-handed quark and
lepton doublets, respectively; $\phi_i$ $(i=1,2)$ are the two Higgs
doublets with ${\widetilde{\phi}}_i = i \tau_2 \phi^*_i $; and $E$,
$F$,$G$ are the $3 \times 3$ matrices in the generation space and
give masses respectively to the charged leptons, the down and up
type quarks; ${\bf 1} ^{(1)} \equiv diag(1,1,0)$; ${\bf 1} ^{(2)}
\equiv diag(0,0,1)$ are the two orthogonal projection operators onto
the first two and the third families respectively. The top quark is
assigned a special status by coupling it to one Higgs doublet that
gets a large VEV, whereas all the other quarks are coupled only to
the other Higgs doublet whose VEV is much smaller. Consequently, if
one  sets the VEVs of $\phi_1$ and $\phi_2$ to be $v_1/\sqrt{2}$ and
$v_2 e^{i \theta} /\sqrt{2}$ \cite{das1996}, the ratio of two Higgs VEVs, $\tan
\beta = v_2 / v_1$, is required to be relatively large.

The Yukawa couplings involving the charged-Higgs bosons are of the form \cite{das1996}
\beq
{\cal L}^C_Y  &=&\frac{g}{\sqrt{2}M_W} \{ - \overline{u}_L V M_D d_R [G^+ - \tan
\beta H^+ ] +\overline{u}_R M_U V d_L [G^+ - \tan \beta H^+] \non
 &&  + \overline{u}_R \Sigma^ {\dag} V d_L [\tan \beta + \cot \beta] H^+ +h.c. \}.
\eeq
where $G^{\pm}$ and $H^{\pm}$ denote the would-be Goldstone bosons
and the physical charged Higgs bosons, respectively. Here $M_U$ and
$M_D$ are the diagonal up- and down-type mass matrices, $V$ is the
usual CKM matrix and $\Sigma \equiv M_U U^{\dag}_R {\bf 1}^{(2)}
U_R$. $U^{\dag}_R$ is the unitary matrix which diagonalizes the
right-handed up-type quarks and has the following form:
\beq
U_R = \left(\begin{array}{ccc} \cos\phi & -\sin\phi & 0 \\
\sin\phi & \cos\phi & 0 \\
 0 & 0 & 1 \end{array} \right) \times \left(\begin{array}{ccc}
1 & 0 & 0 \\  0 & \sqrt{1-|\epsilon_{ct} \xi|^2} & -\epsilon_{ct}
\xi^* \\ 0 & \epsilon_{ct} \xi & \sqrt{1-|\epsilon_{ct} \xi|^2}
\end{array} \right).  \label{eq:ur} \eeq
where $\epsilon_{ct} \equiv m_c/m_t$, $\xi=|\xi|e^{i\delta }$ is
a complex number of order unity, and the phase $\delta$ in $\xi$ is a
new CP violating phase. Inserting Eq.(\ref{eq:ur}) into the definition of
$\Sigma$ yields
\beq
\label{eq:sigma}
\Sigma= \left(\begin{array}{ccc} 0 & 0 & 0 \\
0 & m_c \epsilon_{ct}^2 |\xi|^2 & m_c \epsilon_{ct}
\xi^*\sqrt{1-|\epsilon_{ct} \xi|^2} \\ 0 & m_c \xi
\sqrt{1-|\epsilon_{ct} \xi|^2} & m_t (1-|\epsilon_{ct} \xi|^2)
\end{array} \right).
\eeq

In the following sections, we will calculate the charged Higgs
contributions to the rare decay $\bsll$ in the top quark two-Higgs-doublet model.

\subsection{Effective Hamiltonian for $\bsll$ in the SM }

In the framework of the SM, the effective hamiltonian inducing the
transition $b \rightarrow s l^+ l^-$ at the scale $\mu$ can be
written as follows:
\beq
\label{Heff}
{\cal H} = - \frac{4 G_F}{\sqrt{2}} V^*_{ts} V_{tb} \sum \limits_{i=1}^{10} C_i(\mu) Q_i(\mu),
\eeq
where $G_F$ is the coupling constant, and $V^*_{ts} V_{tb}$ is the
Cabibbo-Kobayashi-Maskawa (CKM) factor \cite{ckm}. The operators can
be chosen as Ref.~\cite{misiak95}:
\begin{align}
Q_1    & =  (\bar{s}_{L}\gamma_{\mu} T^a c_{L }) (\bar{c}_{L }\gamma^{\mu} T^a b_{L}) \, , &
Q_2    & =  (\bar{s}_{L}\gamma_{\mu}  c_{L }) (\bar{c}_{L }\gamma^{\mu} b_{L}) \, , \non
Q_3    & =  (\bar{s}_{L}\gamma_{\mu}  b_{L }) \sum_q (\bar{q}\gamma^{\mu}  q) \, , &
Q_4    & =  (\bar{s}_{L}\gamma_{\mu} T^a b_{L }) \sum_q (\bar{q}\gamma^{\mu} T^a q) \, , \non
Q_5    & =  (\bar{s}_L \gamma_{\mu_1} \gamma_{\mu_2} \gamma_{\mu_3}b_L)
                \sum_q(\bar{q} \gamma^{\mu_1} \gamma^{\mu_2}\gamma^{\mu_3}q)  \, , &
Q_6    & =  (\bar{s}_L \gamma_{\mu_1} \gamma_{\mu_2}  \gamma_{\mu_3} T^a b_L)
                \sum_q(\bar{q} \gamma^{\mu_1} \gamma^{\mu_2}
                \gamma^{\mu_3} T^a q) \, , \non
Q_{7\gamma} & =  \frac{e}{g_s^2} m_b (\bar{s}_{L} \sigma^{\mu\nu} b_{R}) F_{\mu\nu} \, , &
Q_{8g}    & =  \frac{1}{g_s} m_b (\bar{s}_{L} \sigma^{\mu\nu} T^a b_{R}) G_{\mu\nu}^a \, , \non
Q_{9V} & =  \frac{e^2}{g_s^2}(\bar{s}_L\gamma_{\mu} b_L) \sum_\ell(\bar{\ell}\gamma^{\mu}\ell) \, , &
Q_{10A}& =  \frac{e^2}{g_s^2}(\bar{s}_L\gamma_{\mu} b_L)\sum_\ell(\bar{\ell}\gamma^{\mu} \gamma_{5} \ell )
\, , \label{oper}
\end{align}
where $Q_{1,2}$ are the current-current operators, $Q_{3-6}$ the QCD
penguin operators, $Q_{7,8}$ ``magnetic penguin" operators, and
$Q_{9,10}$ semileptonic electroweak penguin operators.
$T^{a}(a=1,...,8)$ stands for $SU(3)_{c}$ generators, $L,R\equiv
(1\mp \gamma_{5})/2$ by definition. The sum over $q$ runs
over the quark fields that are active at the scale $\mu={\cal
O}(m_b)$, i.e., $q\in \left \{ u,d,s,c,b\right \} $. We work in the
approximation where the combination $(V_{us}^* V_{ub})$ of the
Cabibbo-Kobayashi-Maskawa (CKM) matrix elements is neglected. We do
not separate top-quark and charm-quark contributions and will give
the results in the summed form.

To calculate the semileptonic B meson decays at next-to-leading
order in $\alpha_{s}$, we should determinate the Wilson
coefficient $C_{i}(M_{W})$ through matching of the full theory
onto the five-quark low energy effective theory where the $W^\pm$
gauge boson, top quark and the new particles of T2HDM heavier than
$M_{W}$ are integrated out, and run the Wilson coefficients down
to the low energy scale $\mu\sim {\cal O}(m_b)$ by using the QCD
renormalization group equations. The corresponding Wilson
coefficients in SM can be found, for example, in Refs.\cite{ajb98,gam96}.

\begin{figure}[thb]
\vspace{-8cm}
\centerline{\mbox{\epsfxsize=18cm\epsffile{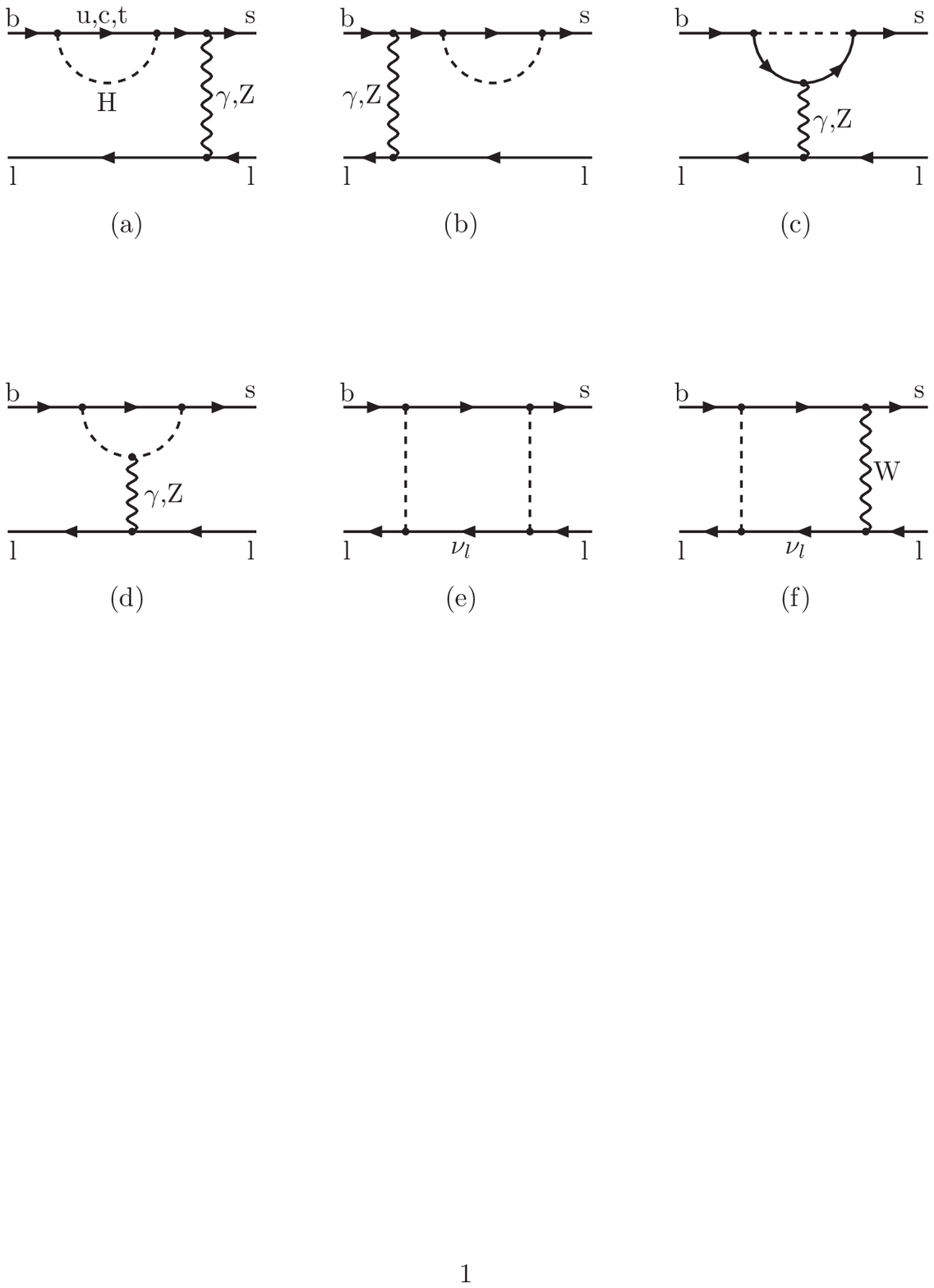}}}
\vspace{-10cm}
\caption{The typical Feynman diagrams for the decay $\bsll$
in the T2HDM. The internal solid and dashed lines denote the
propagators of upper quarks (u,c,t) and charged Higgs boson, respectively.}
\label{fig:fig1}
\end{figure}

\subsection{New physics contributions}

In the framework of the SM, the semileptonic $\bsll$ $(l=e^-,
\mu^-)$ decays proceed through loop diagrams and are of forth order
in the weak coupling. The dominant contributions to this decay come
from the $W$ box and $Z$ penguin diagrams. The corresponding
one-loop diagrams in the SM were evaluated long time ago and can be
found in Refs.~\cite{inami1981,misiak95}. The calculations at the
next-next-to-leading order (NNLO)are also available now.

In the T2HDM considered here, besides the SM diagrams with a W-gauge
boson and an up quark in the loop, the $\bsll$  decays can also
proceed via the new diagrams involving the charged-Higgs boson
exchanges, as illustrated by Fig.\ref{fig:fig1}. In order to
determine the new physics contributions to the relevant Wilson
coefficients $C_{7\gamma}$, $C_{8g}$, $C_{9V}$, and $C_{10A}$ at the
$M_W$ scale, we need to calculate the corresponding Feynman
diagrams.

The new physics parts of the Wilson coefficients $C_{7\gamma}$ and
$C_{8g}$ have been calculated in Refs.~\cite{kiers,kiers62} and
confirmed by our independent calculation. In the naive dimensional
regularization (NDR) scheme, they are of the form
\beq
C_{7\gamma}^{\np}(M_W)&=&\sum \limits_{i=c, t}\kappa^{is}\left[
-\tan^2\beta + \frac{1}{m_i V_{is}^*} (\Sigma^T V^*)_{is}
(\tan^2\beta+1)\right]  \non & & \cdot  \left\{
B(y_i)+\frac{1}{6}A(y_i)\left[-1+ \frac{1}{m_i V_{ib}}
(\Sigma^\dagger V)_{ib}(\cot^2\beta+1)\right]
\right\}, \label{eq:c7gmwnew} \\
C_{8g}^{\np}(M_W)&=& \sum \limits_{i=c, t}\kappa^{is}\left[
-\tan^2\beta + \frac{1}{m_i V_{is}^*} (\Sigma^T V^*)_{is}
(\tan^2\beta+1)\right] \non & & \cdot \left\{
E(y_i)+\frac{1}{6}F(y_i)\left[-1+ \frac{1}{m_i V_{ib}}
(\Sigma^\dagger V)_{ib}(\cot^2\beta+1)\right] \right\},
\label{eq:c8gmwnew}
\eeq
with the Inami-Lim functions
\beq
A(y)&=&\frac{7y -
5y^2-8y^3}{12(1-y)^3}+\frac{2y^2-3y^3}{2(1-y)^4}\ln[y], \non
B(y)&=&\frac{-3y + 5y^2}{12(1-y)^2}-\frac{2y-3y^2}{6(1-y)^3}\ln[y], \non
E(y)&=&\frac{-3y+y^2}{4(1-y)^2}-\frac{y}{2(1-y)^3}\ln[y], \non
F(y)&=&\frac{2y +5y^2-y^3}{4(1-y)^3}+\frac{3y^2}{2(1-y)^4}\ln[y],
\eeq
where $\kappa^{is}=-V_{ib}V_{is}^*/(V_{tb}V_{ts}^*)$, $
y_i=(m_i/m_H)^2$.

As for the Wilson coefficients $C_{9V}$, and $C_{10A}$ at the $M_W$
scale, we found the new physics parts after calculating analytically
the Feynman diagrams as shown in Fig.~\ref{fig:fig1},
\beq
C_{9V}^{\np}(M_W)&=&\frac{1}{\sin^2 \theta_W}
\left[C_0^{\np}-B_0^{\np}\right]
-\left[D_0^{\np}+4 C_0^{\np}\right], \label{eq:c9vmwnew}\\
C_{10A}^{\np}(M_W)&=&-\frac{1}{\sin^2
\theta_W}\left[C_0^{\np}-B_0^{\np}\right], \label{eq:c10amwnew} \eeq
where
\beq B_0^{\np}&=&-\frac{m_l m_b
\tan^2\beta}{8M_W^2}B_{+}(x_{H^+},x_t),
\label{eq:b0np} \\
C_0^{\np}&=&\sum \limits_{i=c, t}\kappa^{is}\frac{m_i^2}{8M_W^2}
\left\{ \left[C_{01}^{'}(y_i)-\frac{4m_b^2}{3m_i^2} \sin^2 \theta_W
C_{11}^{'}(y_i)\right]  \right. \non & & \cdot  \left[ -\tan^2\beta
+ \frac{1}{m_i V_{is}^*} (\Sigma^T V^*)_{is} (\tan^2\beta+1)\right]
\left[-1+ \frac{1}{m_i V_{ib}}(\Sigma^\dagger
V)_{ib}(\cot^2\beta+1)\right] \non & &
+\frac{m_b^2}{m_i^2}\left[(1-\frac{4}{3}\sin^2
\theta_W)C_{01}^{'}(y_i)-C_{11}^{'}(y_i)\right]  \non & & \left.
\cdot \left[ -\tan^2\beta + \frac{1}{m_i V_{is}^*}
(\Sigma^T V^*)_{is} (\tan^2\beta+1)\right] \right\} , \label{eq:b0np2}\\
D_0^{\np}&=& \sum \limits_{i=c, t}\kappa^{is}\frac{2 H(y_i)}{3}
\left[ -\tan^2\beta + \frac{1}{m_i V_{is}^*} (\Sigma^T V^*)_{is}
(\tan^2\beta+1)\right] \non & &\cdot  \left[-1+ \frac{1}{m_i V_{ib}}
(\Sigma^\dagger V)_{ib}(\cot^2\beta+1)\right],\label{eq:d0np} \eeq
with
\beq B_{+}(x,z)&=&\frac{z}{x-z}\left[\frac{\ln[z]}{z-1}- \frac{
\ln[x]}{x-1}\right],\non
H(y)&=&\frac{38y-79y^2+47y^3}{72(1-y)^3}+\frac{4y-6y^2+3y^4}{12(1-y)^4}\ln[y],\non
C_{01}^{'}(y)&=&\frac{y}{1-y}+\frac{y}{(1-y)^2}\ln[y],\non
C_{11}^{'}(y)&=&\frac{3y-y^2}{4(1-y)^2}+\frac{y}{2(1-y)^3}\ln[y].
\eeq
where $ y_i=m_i^2/m_H^2$, $x_{H^+}=m_H^2/M_W^2$, and $x_t
=m_t^2/M_W^2$. $V$ is the CKM matrix, and the matrix $\Sigma$ has
been given in Eq.~(\ref{eq:sigma}). The contributions from
Fig.\ref{fig:fig1}e and the Fig.\ref{fig:fig1}f when the internal W
and charged-Higgs lines exchange their position are strongly
suppressed by a factor of $(m_l/\mh)^2$ ($m_l=m_e, m_\mu$) or
$m_s/m_b$, and therefore have been neglected.

\subsection{The differential decay rate}

Within the Standard Model, the differential decay rate for the decay
$\bsll$ in the NNLO approximation can be written as \cite{buras52,bobeth04}
\beq
\label{eq:RR}
R(\hat{s})& \equiv & \frac{\frac{d}{d\hat{s}}\Gamma(b
\to s l^+ l^-)}{\Gamma(b \rightarrow ce\overline{\nu})}  =
\frac{\alpha_{em}^2}{4 \pi^2}\left|\frac{V^*_{ts}
V_{tb}}{V_{cb}}\right|^2 \frac{  (1-\hat{s} )^2}{f(z)\kappa(z)} \left[
(1+2\hat{s}) \left( \left |\widetilde{C}_{9V}^{eff}(\hat{s})\right |^2 +
\left |\widetilde{C}_{10A}^{eff}(\hat{s}) \right |^2 \right) \right.
\nonumber  \\
& &\left. +4\left(1+\frac{2}{\hat{s}}\right)
\left |\widetilde{C}_{7\gamma}^{eff}\right |^2 + 12 {\rm
Re}\left [\widetilde{C}_{7\gamma}^{eff}\left (\widetilde{C}_{9V}^{eff }(\hat{s}) \right )^* \right ]
\right],
\eeq
where
\beq
\widetilde{C}_k^{eff}= -\widehat{C}_k^{eff} +  \frac{V_{us}^*V_{ub}}{V_{ts}^* V_{tb}}
\delta_{k9} \Delta \widehat{C}_9^{eff}
\eeq
that are related to the evolved coefficients $C_k(\mu_b)$ as
follows:
\beq
\widehat{C}_{7\gamma}^{eff} &=& \f{4 \pi}{\alpha_s(\mu_b)} C_7(\mu_b)
-\f{1}{3} C_3(\mu_b) -\f{4}{9} C_4(\mu_b) -\f{20}{3} C_5(\mu_b)
-\f{80}{9} C_6(\mu_b), \label{c7eff.1}\\
\widehat{C}_{9V}^{eff}(\s) &=& 4 C_9(\mu_b) \left(
\f{\pi}{\alpha_s(\mu_b)} + \omega(\s) \right) + \sum_{i=1}^6
C_i(\mu_b) \gamma^{(0)}_{i9} \ln \f{m_b}{\mu_b} \nonumber\\
&&
+h\left(\f{m_c^2}{m_b^2},\s\right) \left[  \f{4}{3} C_1(\mu_b) +
C_2(\mu_b)
 + 6 C^Q_3(\mu_b) + 60 C^Q_5(\mu_b) \right] \nonumber\\
&&
+ h(1,\s) \left( -\f{7}{2} C_3(\mu_b)-\f{2}{3} C_4(\mu_b)-38
C_5(\mu_b)-\f{32}{3} C_6(\mu_b) \right) \nonumber\\
&&
+  h(0,\s) \left( -\f{1}{2} C_3(\mu_b)-\f{2}{3} C_4(\mu_b)- 8
C_5(\mu_b)-\f{32}{3} C_6(\mu_b) \right) \nonumber\\
&&
+ \f{4}{3} C_3(\mu_b)+ \f{64}{9} C_5(\mu_b)+ \f{64}{27} C_6(\mu_b),
\label{c9eff.1}\\
\widehat{C}_{10A}^{eff}(\s) &=& 4 C_{10}(\mu_b) \left(
\f{\pi}{\alpha_s(\mu_b)} + \omega(\s) \right),\label{c10eff.1}\\
\Delta \widehat{C}_{9V}^{eff} &=& \left[ h(0,\s) -
h\left(\f{m_c^2}{m_b^2},\s\right) \right]
                   \left( \f{4}{3} C_1(\mu_b) + C_2(\mu_b) \right), \label{dc9eff.1}
\eeq
with
\beq
h(z,\s) &=& -\f{4}{9} \ln z +\f{8}{27}+
\f{4}{9}x \non
&& -\f{2}{9}(2+x)\sqrt{|1-x|} \left\{ \begin{array}{ll} \ln
\left|\f{\sqrt{1-x}+1}{\sqrt{1-x}-1}\right|-i\pi,
                               & {\rm for}\;\;x \equiv 4z/\s < 1,\\
2\;{\rm arctan}(1/\sqrt{x-1}), & {\rm for}\;\;x \equiv 4z/\s > 1,\\
\end{array} \right.  \\
h(0,\s)&=& \f{8}{27}-\f{4}{9}(\ln \s- i \pi) , \nonumber \\
 \omega(\s) &=& -\f{4}{3} Li_2(\s) -\f{2}{3}
\ln(1-\s) \ln\s -\f{2}{9} \pi^2
-\f{5+4\s}{3(1+2\s)} \ln(1-\s), \\
&&-\f{2\s(1+\s)(1-2\s)}{3(1-\s)^2(1+2\s)}\ln \s
+\f{5+9\s-6\s^2}{6(1-\s)(1+2\s)},
\eeq
and
\beq
f(z)&=&1-8z^2+8z^6-z^8-24z^4\ln{z}, \label{eq:fz} \\
\kappa(z)&\simeq&1-\frac{2\alpha_s(\mu)}{3\pi} \left[ \left(\pi^2
-\frac{31}{4}\right)(1-z)^2 + \frac{3}{2} \right],
\label{eq:kappaz}
\eeq
here $\hat{s}=(p_{l^+}+p_{l^-})^2/m_b^2 = m_{ll}^2/m_b^2$,
$z=m_c/m_b$, $f(z)$ and $\kappa(z)$ are the
phase-factor and single gluon QCD correction to the $b \to ce\bar{\nu}$ decay, respectively.

In Refs.~\cite{bobeth04}, the Wilson coefficients have been expanded perturbatively as follows
\beq \label{coeffs} C_i = C^{(0)}_i + \frac{g_s^2}{(4 \pi)^2}
C^{(1)}_i + \frac{g_s^4}{(4 \pi)^4} C^{(2)}_i + {\cal O}(g^6).
\eeq
For the standard model parts of the Wilson coefficients $C_i^{(0)}, C_i^{(1)}$ and $C_i^{(2)}$, the explicit
expressions as given in Refs.~\cite{buras52,bobeth04} will be used in our numerical calculation.
For the new physics part, only $C_i^{(1)NP}(\mw)$ are known at present,  as given explicitly in
Eqs.~(\ref{eq:c7gmwnew},\ref{eq:c8gmwnew},\ref{eq:c9vmwnew},\ref{eq:c10amwnew}), and will be included in
numerical calculations.

\section{Numerical result}\label{sec:result}

In this section, we first give the input parameters needed in
numerical calculations, and then present the  numerical results
and make some theoretical analysis.

\subsection{input parameters}

In numerical calculations we will use the following input parameters
(all masses are in GeV) \cite{pdg04}:
\beq M_W&=&80.425,\quad  G_F=1.16639\times 10^{-5} {\rm
GeV}^{-2},\quad \alpha_{em}=1/128, \non
m_c &=&1.4,\quad m_b=4.8 \pm 0.2,
\quad m_t =173.8 \pm 5,\quad \Lambda^{(5)}_{\overline{\rm
MS}}=0.225,  \non A&=&0.853,\quad \lambda=0.2200, \quad \rho=0.20
\pm 0.09,\quad \eta=0.33\pm 0.05, \non \quad \ssa &=&0.23124, \quad
Br(B \to X_c e \bar{\nu})=0.1061, \label{eq:input}
\eeq where the
parameter $A, \lambda, \rho$ and $\eta$ are Wolfenstein parameters
of the CKM mixing matrix. For the strong coupling constant
$\alpha_s(\mu)$ we use the two-loop expression,
\beq
\alpha_s(\mu) = \frac{4\pi}{\beta_0 \,
\ln(\mu^2/\Lambda_{\overline{\rm MS}}^2)}\; \left[ 1 -
\frac{\beta_1}{\beta_0^2}\cdot \frac{\ln
\ln(\mu^2/\Lambda_{\overline{\rm
MS}}^2)}{\ln(\mu^2/\Lambda_{\overline{\rm MS}}^2)} \right],
\eeq
with
\beq \beta_0 = \frac{33-2f}{3}, \ \ \beta_1 = 72 -10 f -8f/3
\eeq
where the $f$ is the number of quark flavors and the term $\overline{\rm MS}$ denotes the
modified subtraction scheme.

\subsection{$B \to X_s \gamma$ decay}

There are four free parameters $m_H$, $\tan\beta$, $|\xi|$ and a new
CP-violating phase $\delta$ in the T2HDM. We fix $|\xi|=1$ throughout the paper and
consider other three as variable parameters to be constrained by precise measurements,
such as the date of $Br(B \to X_s \gamma)$.

\begin{figure}[thb]
\centerline{\mbox{\epsfxsize=10cm\epsffile{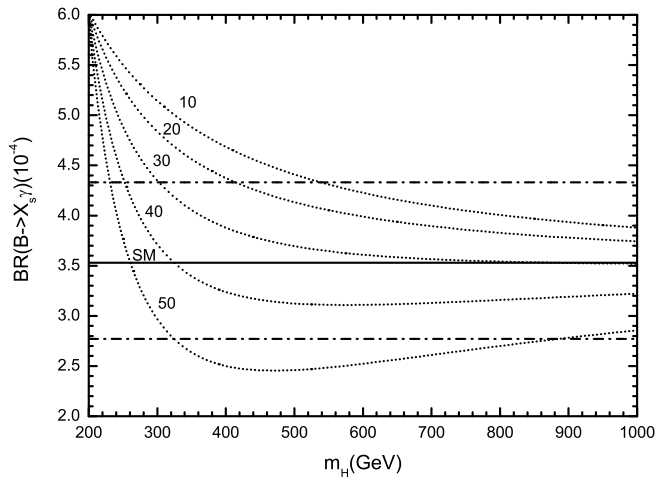} }}
\vspace{0.2cm}
\caption{The $m_H$ dependence of $Br(B \to X_s \gamma)$ in the T2HDM for
$\delta=0^\circ$, and for $\tan\beta=10$, $20$, $30$, $40$ and $50$ respectively.
The band between two horizontal dash dot lines shows data as specified in Eq.(\ref{eq:bsgdata}).
The solid horizontal line shows the central value of the SM prediction.}
\label{fig:fig2}
\end{figure}

In Ref.~\cite{xiao05}, the branching ratio $Br(B \to X_s \gamma)$
have been calculated in both the SM and the T2HDM. Using the
formulas as given in Appendix \ref{app:aa} and  taking the range of
\beq
 2.77 \times 10^{-4} \leq  Br(B \to  X_s \gamma)
\leq  4.33 \times 10^{-4} \label{eq:bsgdata}
\eeq
as the experimentally allowed region at $3\sigma$ level \cite{hfag06},
one can read off the lower limit on the mass of charged-Higgs boson $m_H$ directly
from Fig.~\ref{fig:fig2}:
\beq
m_H \geq  300{\rm GeV}, \label{eq:mhbound}
\eeq
for fixed $\tan{\beta} =30$ and $\delta=0^\circ$.
It is easy to see from Fig.~\ref{fig:fig2} that
(a) a light charged Higgs boson with a mass less than 200 GeV is excluded by the data of
$B \to X_s \gamma$ decay at $3\sigma$ level; and (b) a charged-Higgs boson with a mass
heavier than $300$ GeV is still allowed by the same data.

\begin{figure}[thb]
\centerline{\mbox{\epsfxsize=10cm\epsffile{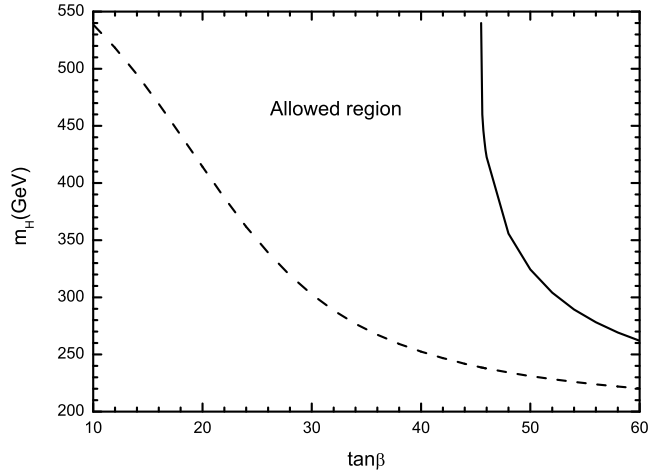}}}
\vspace{0.2cm}
\caption{Contour plot in $\tan\beta-m_H$ plane obtained by considering the data
in Eq.~(\ref{eq:bsgdata}) for fixed $\delta=0$.
The region between the short-dashed and solid curves is still allowed by the data
as given in Eq.~\ref{eq:bsgdata}. }
\label{fig:fig3}
\end{figure}

As shown in the contour plot Fig.~\ref{fig:fig3}, the region between the short-dashed and
solid curves is still allowed by the data of $B \to X_s \gamma$ as given in Eq.~(\ref{eq:bsgdata})
for fixed value of $\delta=0^\circ$.
On the other hand, by assuming $\tan\beta=30$ and $m_H=400$ GeV, one finds a strong
constraint on the phase $\delta$: $\delta < 44^\circ$.

\subsection{$B\to X_s l^+ l^-$ decay }

The branching ratio of $B \to X_s l^+ l^-$ $(l=e,\mu)$ has been
recently measured by BaBar and Belle
Collaborations\cite{babar-04a,belle-05a}. In the low-$q^2$ region
\footnote{The low-$q^2$ region is the region with $1{\rm GeV^2} \leq
m_{ll}^2 \equiv q^2 \leq 6{\rm GeV^2}$.}, the average of BaBar and
Belle's measurements is \cite{gambino05}
\beq
Br(B \to  X_s l^+ l^-) = (1.60 \pm 0.51) \times 10^{-6}.
\label{eq:average}
\eeq

Theoretically, the integrated branching ratio can be written as
\cite{bobeth04}
\beq
Br_{ll} = Br(\bar{B} \to X_c l \nu)
\int_{\hat{s}_a}^{\hat{s}_b} \; R(\hat{s}),
\label{eq:bsll}
\eeq
where $\hat{s} = q^2/m_b^2$ with $\hat{s}_a =1/m_b^2$ and $\hat{s}_b
=6/m_b^2$, and the differential decay rate $R(\hat{s})$ has been
defined in Eq.~(\ref{eq:RR}). The SM prediction after integrating
over the low-$q^2$ region reads
\beq Br(B \to X_s l^+ l^-) &=& \left
( 1.58 \pm 0.08|_{m_t} \pm 0.07|_{\mu_b} \pm 0.04|_{CKM} \pm
0.06|_{m_b} + 0.18|_{\mu_w}  \right ) \times 10^{-6}  \non
&=&\left
( 1.58 \pm 0.13 + 0.18|_{\mu_W}  \right ) \times 10^{-6}.
\label{eq:smp}
\eeq
where the errors correspond to the uncertainty
of input parameters of $m_t$, $A$, $\rho$, $\eta$ and $m_b$ as shown
in Eq.~(\ref{eq:input}), and for $m_b/2 \leq \mu_b \leq 2 m_b$. The
last error refers to the choice of $\mu_W =120$ GeV, instead of
$\mu_W =\mw$. Since we here focus on the new physics contributions
to the branching ratios of $B \to X_s l^+ l^-$ decay, we will take
$\mu_W =\mw$ in the following without further specification.

\begin{table}[t]
\begin{center}
\caption{The effective Wilson coefficients and the interference term ($ 12 Re[\widetilde{C}_{7\gamma}^{eff}
(\widetilde{C}_{9V}^{eff})^*]$) for fixed $\hat{s}=q^2/m_b^2=0.2$, the branching ratio integrated
over the low-$q^2$ region in units $10^{-6}$ in the SM and the T2HDM for $\mh=300$,
 $\tan{\beta}=10,30,50$ and $\delta=0^\circ$ (a), $30^\circ$ (b) and $60^\circ$ (c).
 Only the central values are shown here.}
\label{tab:brlla}
\vspace{0.2cm}
\begin{tabular}{ l|l|l|l|l|l } \hline\hline
                   & $\widetilde{C}_{7\gamma}^{eff}$ & $\widetilde{C}_{9V}^{eff}$ &
                   $\widetilde{C}_{10A}^{eff}$ &Int. Term & $Br_{ll}$\\ \hline
SM                   &    $-0.344$         &     $4.302+ i 0.064$ &    $-3.547          $& $ -17.73$ & $ 1.579$ \\ \hline
T2HDM                &(a) $-0.422+i 0.001$ &(a)  $4.205+ i 0.063$ &(a) $-3.552          $& $ -21.30$ & $ 1.576$\\
$\tan{\beta}=10$     &(b) $-0.424+i 0.006$ &(b)  $4.218+ i 0.014$ &(b) $-3.552+ i 0.001 $& $ -21.45$ & $ 1.581$\\
                     &(c) $-0.428+i 0.010$ &(c)  $4.255- i 0.021$ &(c) $-3.553+ i 0.001 $& $ -21.84$ & $ 1.595$\\ \hline
                     &(a) $-0.376+i 0.001$ &(a)  $3.430+ i 0.051$ &(a) $-3.546          $& $ -15.47$ & $ 1.342$\\
$\tan{\beta}=30$     &(b) $-0.389+i 0.050$ &(b)  $3.554- i 0.385$ &(b) $-3.546+ i 0.001 $& $ -16.84$ & $ 1.388$\\
                     &(c) $-0.425+i 0.086$ &(c)  $3.879- i 0.700$ &(c) $-3.547+ i 0.002 $& $ -20.50$ & $ 1.581$\\ \hline
                     &(a) $-0.283+i 0.002$ &(a)  $1.882+ i 0.026$ &(a) $-3.544          $& $  -6.40$ & $ 1.033$\\
$\tan{\beta}=50$     &(b) $-0.321+i 0.140$ &(b)  $2.226- i 1.183$ &(b) $-3.544+ i 0.002 $& $ -10.56$ & $ 1.167$\\
                     &(c) $-0.420+i 0.239$ &(c)  $3.126- i 2.056$ &(c) $-3.546+ i 0.003 $& $ -21.65$ & $ 1.526$\\ \hline
\hline
\end{tabular}\end{center}
\end{table}

Now we consider the new physics contributions. When the new physics
parts of the Wilson coefficients $C_i^{(1)(\mw)}$ for $i=7\gamma,
8g, 9V$ and $10A$ are taken into account, the values of the
effective Wilson coefficients appeared in Eq.~(\ref{eq:RR}) and the
theoretical predictions of the branching ratio will be changed
accordingly, as listed in Table \ref{tab:brlla} for
$\tan{\beta}=10,30,50$, $\mh=300$ GeV and $\delta=0^\circ, 30^\circ$
and $60^\circ$ .

\begin{figure}[thb]
\centerline{\mbox{\epsfxsize=10cm\epsffile{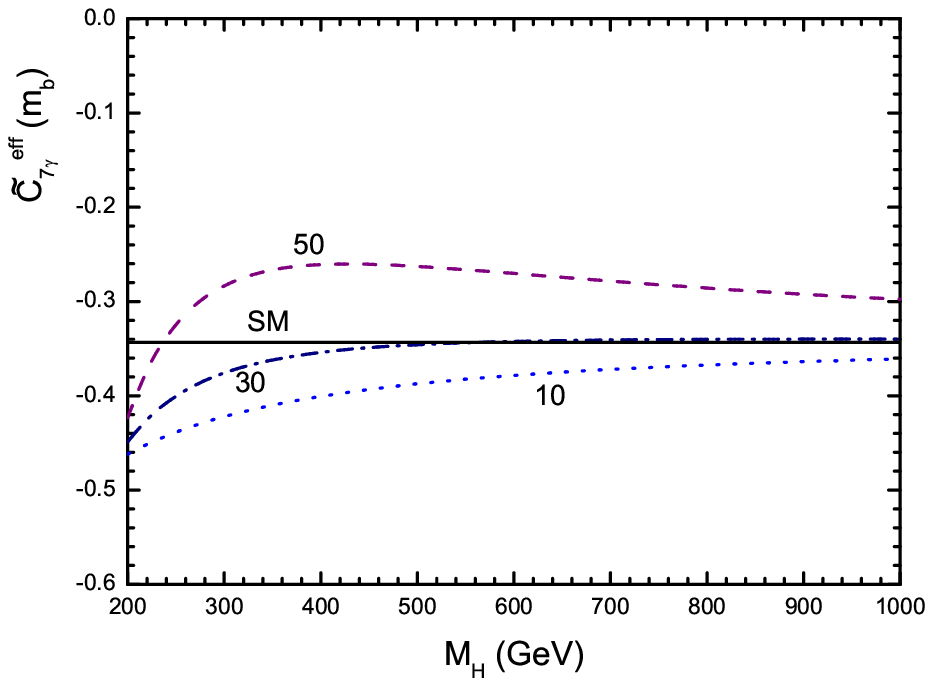} }}
\caption{The $\mh$ dependence of the real part of the effective
Wilson coefficient $\widetilde{C}_{7\gamma}^{eff}(m_b)$ in the SM (solid line) and T2HDM
for $\delta=0^\circ$, and $\tan\beta=10$ (dots curve), $30$ (dot-dashed curve) and $50$
(dashed curve), respectively. }
\label{fig:fig4}
\end{figure}

\begin{figure}[tb]
\vspace{-1cm}
\centerline{\mbox{\epsfxsize=10cm\epsffile{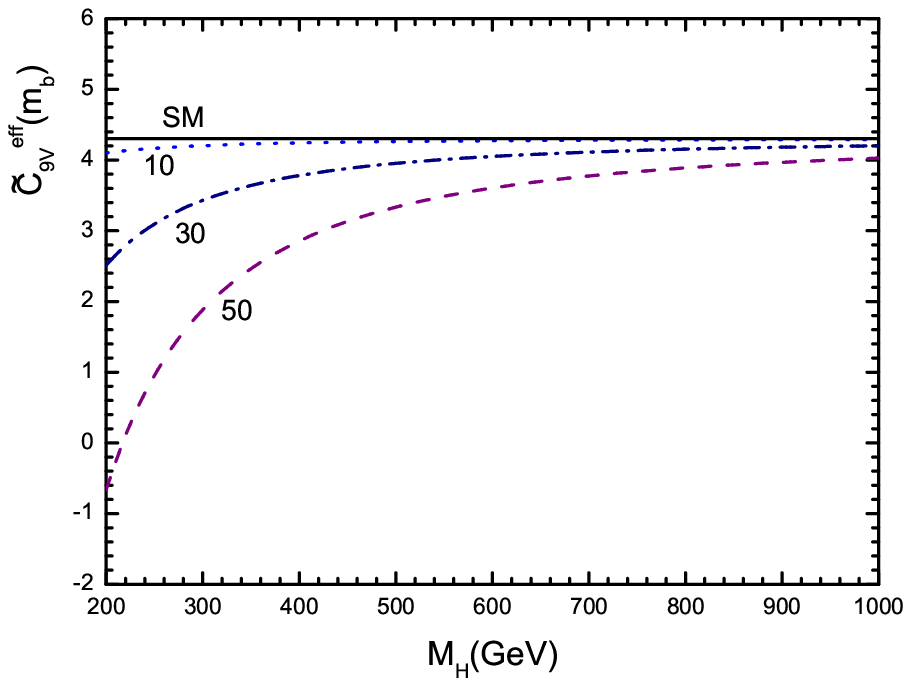} }}
\caption{The $\mh$ dependence of the real part of the effective Wilson coefficient
$\widetilde{C}_{9V}^{eff}(m_b)$ in the SM (solid line) and T2HDM
for $\hat{s}=0.2$, $\delta=0^\circ $, and $\tan\beta=10$ (dots curve), $30$ (dot-dashed curve) and $50$
(dashed curve), respectively. }
\label{fig:fig5}
\end{figure}

\begin{figure}[tb]
\centerline{\mbox{\epsfxsize=10cm\epsffile{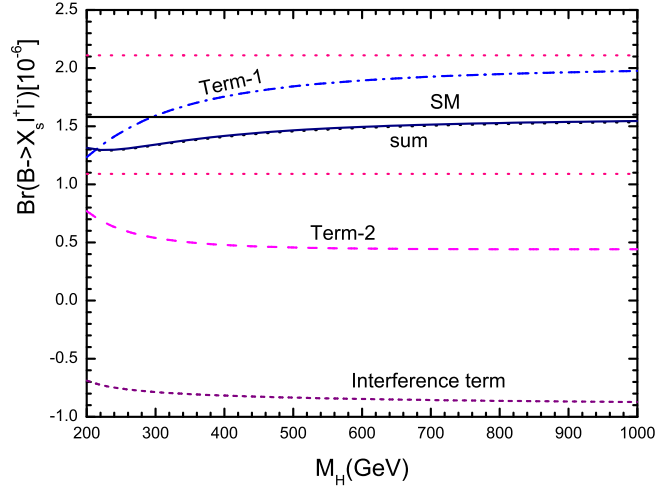} }} \caption{The
$m_H$ dependence of the branching ratio of $B \to X_s l^+ l^-$ in
the SM and T2HDM for $\delta=0^\circ$, and $\tan{\beta}=30$. The
contributions from the term-1, term-2, interference term and their
summation  are shown by the dot-dashed, dashed, short-dashed and
solid curve, respectively. The horizontal band between two dots line
shows the data: $Br(B \to X_s l^+ l^-)=(1.60 \pm 0.51) \times
10^{-6}$, while the solid line refers to the central value of SM
prediction: $Br(B \to X_s l^+ l^-)=1.58\times 10^{-6}$.}
\label{fig:fig6}
\end{figure}

\begin{figure}[tb]
\centerline{\mbox{\epsfxsize=10cm\epsffile{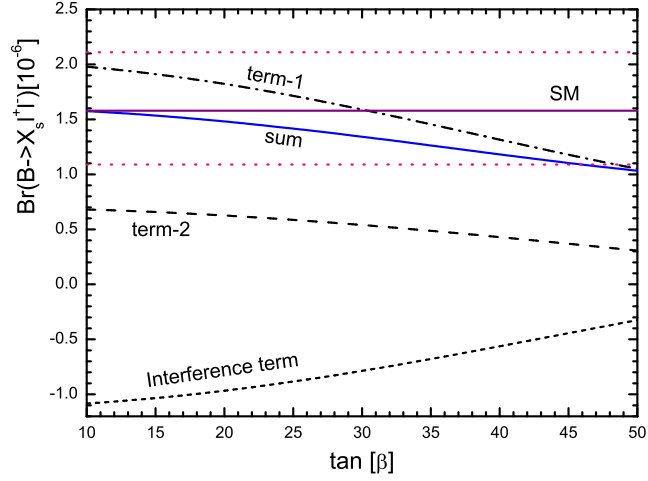} }}
\caption{The same as Fig.~\ref{fig:fig6}, but shows the $\tan{\beta}$ dependence of the branching ratio
of $B \to X_s l^+ l^-$ in the SM and T2HDM for $\delta=0^\circ$ and $\mh=300$ GeV.}
\label{fig:fig7}
\end{figure}

\begin{figure}[tb]
\centerline{\mbox{\epsfxsize=10cm\epsffile{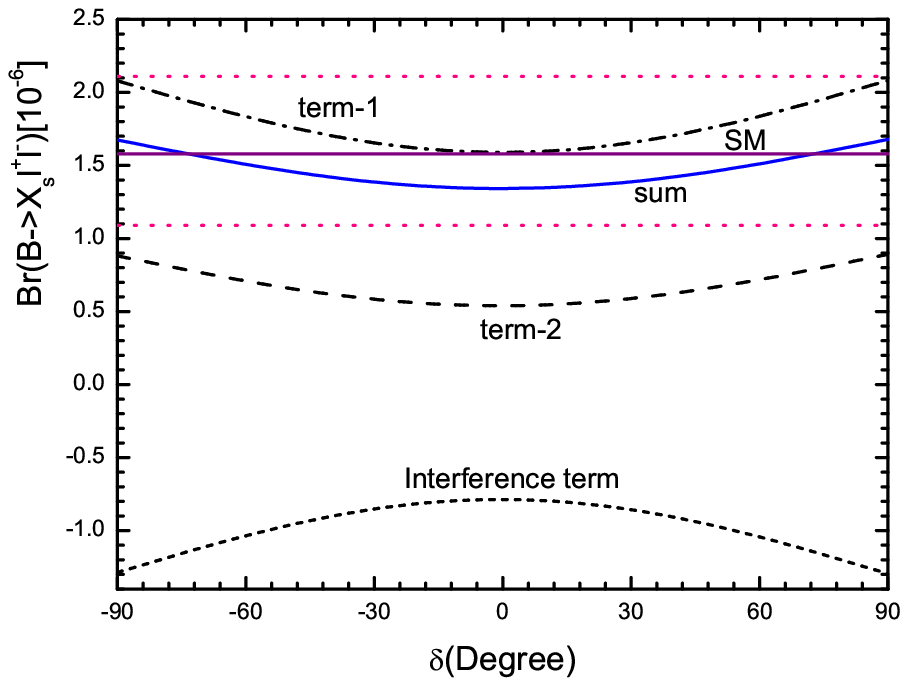} }}
\vspace{0.2cm}
\caption{The same as Fig.~\ref{fig:fig6}, but shows the $\delta$ dependence of the branching ratio
of $B \to X_s l^+ l^-$ in the SM and T2HDM for $\tan{\beta}=30$ and $\mh=300$ GeV.}
\label{fig:fig8}
\end{figure}

In Figs.~\ref{fig:fig4} and \ref{fig:fig5}, in order to show more details of the $\mh$ and $\tan{\beta}$
dependence,  we draw the real part of the effective Wilson coefficients $\widetilde{C}_{7\gamma}^{eff}(m_b)$
and $\widetilde{C}_{9V}^{eff}(m_b)$ for fixed $\hat{s}=q^2/m_b^2=0.2$ and $\delta=0^\circ$.
Within the considered parameter space of the T2HDM, it is easy to see from the numerical results in
Table \ref{tab:brlla} and Figs.~\ref{fig:fig4} and \ref{fig:fig5} that

\begin{enumerate}
\item[]{(i)}
The effective Wilson coefficient $\widetilde{C}_{7\gamma}^{eff}(m_b)$ is always SM-like. This feature
can be seen explicitly in Fig.~\ref{fig:fig4}, where the $\mh$-dependence of the real part
of $\widetilde{C}_{7\gamma}^{eff}(m_b)$ is shown for $\delta=0^\circ$, $\tan{\beta}=10,30, 50$ and
$200 {\rm GeV} \leq \mh \leq 1000 {\rm GeV}$. The imaginary part of $\widetilde{C}_{7\gamma}^{eff}(m_b)$
is generally small.

\item[]{(ii)}
The effective Wilson coefficient $\widetilde{C}_{9V}^{eff}(m_b)$ is also SM-like.
The imaginary part of $\widetilde{C}_{9V}^{eff}(m_b)$ is also generally small.

\item[]{(iii)}
The new physics contribution to $\widetilde{C}_{10A}^{eff}$ is very small in size, less than $1\%$
of its standard model counterpart, and therefore can be neglected safely.

\item[]{(iv)}
The new physics contributions to $\widetilde{C}_{7\gamma}^{eff}$ and $\widetilde{C}_{9V}^{eff}$
can be significant in magnitude respectively for large $\tan{\beta}$, large $\delta$ and lighter
charged-Higgs boson, as can be seen from the numerical results in Table \ref{tab:brlla} and
illustrated explicitly by Figs.\ref{fig:fig4} and \ref{fig:fig5}.
But they tend to cancel each other
and finally lead to a small change to the prediction for the branching ratio under study.

\end{enumerate}
It is worth noting that both the real and imaginary parts of effective Wilson coefficients
are taken into account in our calculation of the branching ratio.

As shown in Eq.~(\ref{eq:RR}), the differential decay rate depends
on the summation  of three terms: \beq Term-1: \quad && (1+
2\hat{s})\;\left (\left |\widetilde{C}_{9V}^{eff}(\hat{s}) \right
|^2 + \left |\widetilde{C}_{10A}^{eff}(\hat{s})\right |^2 \right ),
\non Term-2: \quad && 4\left(1+\frac{2}{\hat{s}}\right)\left
|\widetilde{C}_{7\gamma}^{eff}\right |^2, \non Term-3: \quad && 12
\; {\rm Re}\left [\tilde{C}_{7\gamma}^{eff} \left
(\tilde{C}_{9V}^{eff }(\hat{s})\right )^* \right ], \eeq where the
third term is the interference term, which has opposite sign
compared to first two terms. From Fig.~\ref{fig:fig6}, one can see
easily that
\begin{itemize}
\item[]{(i)}
After the inclusion of new physics contributions in T2HDM, the signs
of three terms remain unchanged.

\item[]{(ii)}
The new physics contributions to these three terms are indeed tend
to cancel each other and result in a  summation (solid curve in
Fig.~\ref{fig:fig6}) which becomes closer to the SM prediction
(solid line in Fig.~\ref{fig:fig6}) when $\mh$ becoming larger. The
theoretical predictions for the branching ratio in the SM and T2HDM
agree well for the whole range of $\mh$ considered here. They are
also in good agreement with the data within one standard deviation.
\end{itemize}

Analogous to Fig.~\ref{fig:fig6}, the Figs.~\ref{fig:fig7} and \ref{fig:fig8} show the
$\tan{\beta}$ and $\delta-$dependence of the branching ratio $Br(B \to X_s l^+ l^-)$, respectively.
Here, the
cancelation of new physics contributions to different terms occurs and leaves the summation,
the theoretical prediction in the T2HDM, in good agreement with the SM prediction
as well as the measured value within one standard deviation.
From  Fig.~\ref{fig:fig7}, one can also see that a $\tan{\beta}$ smaller than $40$ is preferred
by current data.

\section{Summary}

In this paper, we calculate the new physics contributions to the branching ratio of
$B \to X_s \gamma$ and $B \to X_s l^+ l^-$ decays induced by the charged Higgs loop
diagrams in the top-quark two-Higgs-doublet model, and compare the theoretical predictions
in the T2HDM with currently available data.

In Sec.~\ref{sec:th}, we firstly present a brief review about the
basic structure of the top-quark two-Higgs-doublet model, and then
evaluate analytically the new Feynman diagrams induced by the
charged Higgs $H^{\pm}$ exchanges and extract the new physics parts
of the Wilson coefficients $C_{7\gamma}^{\np}(\mu_W)$,
$C_{8g}^{\np}(\mu_W)$, $C_{9V}^{\np}(\mu_W)$ and
$C_{10A}^{\np}(\mu_W)$ which govern the new physics contributions to
$B \to X_s \gamma$ and $B \to X_s l^+ l^-$ decays considered in this
paper. For the SM part, we use the known analytical formulae at NNLO
level as given for example in Refs.~\cite{buras52,bobeth04}. The new
physics contributions are included through the modifications to the
corresponding Wilson coefficients at matching scale $\mu_W \sim
\mw$.

From the numerical results and the figures as shown in Sec.~\ref{sec:result}, we  found that
\begin{enumerate}

\item[]{(i)}
For the T2HDM studied here, a light charged Higgs boson with a mass
less than 200 GeV is excluded by the data of $B \to X_s \gamma$
decay at $3\sigma$ level. But a charged-Higgs boson with a mass
heavier than $300$ GeV is still allowed by the data of
both $B \to X_s \gamma$ and $B \to X_s l^+ l^-$ decay. The data of
$B \to X_s \gamma$ also prefer a small $\delta$, a new CP violating phase
appeared in the Yukawa couplings of the T2HDM.

\item[]{(ii)}
After the inclusion of new physics contributions, the effective
Wilson coefficients $\widetilde{C}_{i}^{eff}(m_b)$ ($i =7\gamma, 9V$
and $10A$), which govern the branching ratio of $B \to X_s l^+ l^-$
decay, are always SM-like within the considered parameter space of
T2HDM. The sign of the interference term in Eq.~(\ref{eq:RR})
remains unchanged.

\item[]{(iii)}
The new physics contributions to $\widetilde{C}_{7\gamma}^{eff}$ and $\widetilde{C}_{9V}^{eff}$
can be significant in magnitude respectively for large $\tan{\beta}$, large $\delta$ and lighter
charged-Higgs boson, but they tend to cancel each other and finally result in only a small change
to the prediction for the branching ratio of $B \to X_s l^+ l^-$ decay. This feature can be seen clearly
through the numerical results in Table \ref{tab:brlla} and the curves shown in last three figures.

\item[]{(iv)}
Within the considered parameter space of the T2HDM, the T2HDM predictions for $Br(B \to X_s l^+ l^-)$
agree well with the SM as well as the measured value within one standard deviation.

\end{enumerate}

\begin{acknowledgments}

This work is partly supported  by the National Natural Science Foundation of China under Grant
No.10275035,10575052, and by the Specialized Research Fund for the Doctoral Program of higher
education (SRFDP) under Grant No.~20050319008.

\end{acknowledgments}


\begin{appendix}

\section{$Br(B \to X_s \gamma)$ in the SM and T2HDM}\label{app:aa}

The branching ratio of ${B\to X_s \gamma}$ at the next-to-leading order (NLO) in the SM and the leading
order (LO) in the T2HDM  can be written as \cite{bg98,xiao05}
\beq
{\cal B}(B \to X_s \gamma) ={ \cal B}_{SL}
\left |\frac{V_{ts}^*V_{tb}}{V_{cb}} \right |^2
\frac{6\alpha_{em}}{\pi f(z) \kappa (z)}[|\bar{D}|^2 + A + \triangle],
\label{eq:br-sm1}
\eeq
where ${\cal B}_{SL}=10.61\%$ is the measured semileptonic branching ratio
of B meson,  $\alpha_{em}$=1/128 is the fine-structure constant, $z=m_c^{pole}/m_b^{pole}=0.29\pm 0.02$ is
the ratio of the quark pole mass. The function $f(z)$ and $\kappa(z)$ have been given in Eqs.~(\ref{eq:fz})
and (\ref{eq:kappaz}).

The term $\bar{D}$ at low energy scale $\mu = {\cal O}(m_b) $ in Eq.~(\ref{eq:br-sm1})
corresponds to the subprocess  $b\to s\gamma$
\beq
\bar{D}=C_{7\gamma}^{\sm}(\mu) + V(\mu) + C_{7\gamma}^{\np}(\mu).
\eeq
Here $C_{7\gamma}^\sm(\mu)$ denotes the SM part of the Wilson coefficient $C_{7\gamma}(\mu)$ at NLO level,
and the explicit expression of $C_{7\gamma}^\sm(\mu)$ at both LO and NLO level can be found easily in
Ref.~\cite{gam96}.

The new physics part of the Wilson coefficient  $C_{7\gamma}$ and  $C_{8g}$ at
the matching scale $M_W$ are currently known at LO level and have been given in
Eqs.~(\ref{eq:c7gmwnew}) and (\ref{eq:c8gmwnew}).
At the low energy scale $\mu = {\cal O}(m_b)$, the leading order
Wilson coefficients $C_{7\gamma}^{\np}(\mu)$  and  $C_{8g}^{\np}(\mu)$ can be written as
\beq
C_{7\gamma}^\np(\mu)  &=&  \eta^\frac{16}{23} C_{7\gamma}^\np(\mw) +
\frac{8}{3} \left (\eta^\frac{14}{23} -\eta^\frac{16}{23}\right ) \; C_{8g}^\np(\mw),
\label{eq:c7npmb}\\
 C_{8g}^\np (\mu) & = &   \eta^\frac{14}{23}  \; C_{8g}^\np(\mw),
 \label{eq:c8gnpmb}
\eeq
where $\eta=\alpha_s(\mw)/\alpha_s(\mu)$, and the Wilson coefficient $ C_{8g}^\np(\mw) $ has been given
in Eq.~(\ref{eq:c8gmwnew}).

The function $V(\mu)$ in Eq.~(\ref{eq:br-sm1}) is defined as \cite{bg98}
\beq
 V(\mu)=\frac{\alpha_s(\mu)}{4 \pi}\left \{  \sum_{i=1}^8 C_i^{0}(\mu)
   \left[r_i+ \frac{1}{2}\gamma_{i7}^{0}
             \ln \frac{m^2_b}{\mu^2}    \right]
             - \frac{16}{3} C_{7\gamma}^{0}(\mu) \right \} \,,
             \label{eq:vmub}
\eeq
where the functions $r_i$ $(i=1,\ldots, 8)$ are the virtual correction functions (see
Appendix D of Ref.~\cite{bg98}), $\gamma_{i7}^{0}$ are the elements
of the anomalous dimension matrix which govern the evolution of the Wilson
coefficients from the matching scale $\mw$ to lower scale $\mu$. The values of
$\gamma_{i7}^{0}$ can be found in Ref.~\cite{bg98}.

In Eq.(\ref{eq:br-sm1}), the term $A=A(\mu)$ is the the correction coming from the
bremsstrahlung process $b \to s\gamma g$ \cite{ag91}
\beq
A(\mu) = \frac{\alpha_s(\mu)}{\pi} \sum_{i,j=1;i \le j}^8 \,
  {\rm Re} \left\{ C_i^{0}(\mu) \,
            \left[ C_j^{0}(\mu)\right]^*
 \, f_{ij} \right\} \, \label{eq:aa}.
\eeq
The coefficients $f_{ij}$ have been defined and computed in Refs.\cite{ag91,cmm97}. We
here use the explicit expressions of those relevant $f_{ij}$ as given in Appendix E of
Ref.\cite{bg98}.

Finally, the term $\Delta $ in Eq.(\ref{eq:br-sm1}) denotes the non-perturbative corrections
\cite{falk94,voloshin97},
\beq
\Delta = \frac{\delta_\gamma^{\np}}{m_b^2} \left |C_7^{0}(\mu)\right |^2
+ \frac{\delta_c^{\np}}{m_c^2} {\rm Re}\left \{ \left [ C_7^{0}(\mu)\right ]^*
\cdot \left [ C_2^{0}(\mu)-\frac{1}{6}C_1^{0}(\mu) \right ] \right \}
\eeq
with
\beq
\delta_{\gamma}^{\np} = \frac{\lambda_1}{2} - \frac{9}{2}\lambda_2, \quad
\delta_{c}^{\np} = - \frac{\lambda_2}{9},
\label{eq:dcnp}
\eeq
where $\lambda_2 = (m_{B^*}^2 - m_B^2)/4 = 0.12$ GeV$^2$ and $\lambda_1=0.5\, {\rm GeV}^2$.

In the expressions of $ V(\mu)$, $A(\mu)$ and $\Delta$, the superscript $``0"$
means that the corresponding Wilson coefficients at LO level will be used.
The numerical results show that the new physics contributions
to ``small quantities" $A(\mu)$ and $\Delta$ are very small in magnitude and can be
neglected safely.

\end{appendix}



\begin{thebibliography}{99}

\bibitem{hfag06}
Heavy Flavor Averaging Group, E.~Barberio  {\it et al}., hep-ex/0603003.

\bibitem{gambino}
K.~Chetyrkin, M.~Misiak, and M.~M\"unz, \plb, {\bf 400}, 206 (1997), {\bf 425}, 414(E) (1998);
P.~Gambino and M.~Misiak, \npb {\bf 611}, 338 (2001);
A.J.~Buras, A.~Czarnecki, M.~Misiak, and J.~Urban, \npb {\bf 631},  219 (2002);
A.J.~Buras, A.~Poschenrieder, M.~Spranger, A.~Weiler, \npb {\bf 678}, 455 (2004).

\bibitem{gambino05}
P.~Gambino, U.~Haisch, and M.~Misiak, \prl {\bf 94}, 061803 (2005).

\bibitem{misiak95}
M.~Misiak, \npb {\bf 393}, 23 (1993); \npb {\bf 439}, 461(E) (1995);
B.~Grinstein, M.J.~Savage, and M.B.~Wise, \npb {\bf 319}, 271 (1989);
R.~Grigjanis, P.J.~O'Donnell, M.~Sutherland, and H.~Navelet, \plb {\bf 223}, 239 (1989).

\bibitem{buras52}
A.J.~Buras, M.~M\"unz, \prd {\bf 52},186 (1995).

\bibitem{huang}
Y.B.~Dai, C.S.~Huang, and H.W.~Huang, \plb {\bf 390}, 257(1997)£»
S.~Schilling, C.~Greub, N.Salzmann and B.T\"odtli, \plb {\bf 616}, 93  (2005).

\bibitem{ali02}
A.~Ali, E.~Lunghi, C.~Greub, and G.~Hiller, \prd {\bf 66}, 034002 (2002).

\bibitem{bobeth05}
C.~Bobeth, A.J.~Buras, F.~Fr\"uger and J.~Urban, \npb {\bf 630}, 87 (2002);
C.~Bobeth, A.J.~Buras and T.~Ewerth, \npb {\bf 713}, 522 (2005).

\bibitem{das1996}
A.~Das and C.~Kao, \plb {\bf 372}, 106 (1996).

\bibitem{kiers}
K.~Kiers, A.~Soni and G.H.~Wu, \prd {\bf 59}, 096001 (1999);
G.H.~Wu and A.~Soni, \prd {\bf 62}, 056005 (2000).

\bibitem{kiers62}
K.~Kiers, A.~Soni and G-H.~Wu, \prd, {\bf 62}, 116004(2000).

\bibitem{hou1992}
W.-S.~Hou, \plb {\bf 296}, 179 (1992);
M.~Luke and M.J.~Savage, \plb {\bf 307}, 387 (1993);
D.~Atwood, L.~Reina and A.~Soni, \prd {\bf 55}, 3156 (1997);
M.~Sher, Proceedings of the 29th International Conference
on High-Energy Physics (ICHEP98), Vancouver, Cananda, 1998, hep-ph/9809590.

\bibitem{ckm}
M.~Kobayashi, T.~Maskawa, Prog. Theor. Phys, {\bf 49}, 652 (1973).

\bibitem{ajb98}
A.J.~Buras, in {\em Probing the Standard Model of Particle
Interactions}, F.David and R. Gupta, eds., 1998, Elsevier Science B.V.; hep-ph/9806471.

\bibitem{gam96}
G.~Buchalla, A.J.~Buras and M.E.~Lautenbacher, \rmp {\bf 68}, 1125 (1996).

\bibitem{inami1981}
T.~Inami and C.S.~Lim,  Prog. Theor. Phys, {\bf 65}, 297 (1981) [erratum {\bf 65}, 1772 (1981)].

\bibitem{bobeth04}
C.~Bobeth, M.~Misiak, and J. Urban \npb {\bf 574}, 291 (2000);
C.~Bobeth, P.~Gambino, M.~Gorbahn and U.~Haisch, JHEP {\bf 0404}, 071 (2004).

\bibitem{pdg04}
Particle Data Group, S.~Eidelman {\it et al.}, \plb {\bf 592}, 1 (2004).

\bibitem{xiao05}
Z.J.~Xiao, H.H.~Cheng and L.X.~L\"u, hep-ph/0512359.

\bibitem{babar-04a}
BaBar Collaboration, B.~Aubert {\it et al}., \prl {\bf 93}, 081802 (2004).

\bibitem{belle-05a}
Belle Collaboration, M.~Iwasaki {\it et al}., \prd {\bf 72} , 092005 (2005);
Belle Collaboration, K.~Abe {\it et al}., hep-ex/0508009.

\bibitem{bg98}
F.M.~Borzumati and C.~Greub, \prd{\bf 58}, 074004 (1998); {\it ibid } {\bf 59}, 057501 (1999).

\bibitem{ag91}
A.~Ali and C.~Greub, Z. Phys. C {\bf 49}, 431 (1991); \plb {\bf 259}, 182 (1991);
{\bf 361}, 146 (1995);  N.~Pott, \prd {\bf 54}, 938 (1996).

\bibitem{cmm97}
K.G.~Chetyrkin, M.~Misiak, M.~Munz, \plb {\bf 400}, 206(1997), {\bf 425}, 414(E) (1998).

\bibitem{falk94}
A.F.~Falk, M.~Luke  and M.J.~Savage, \prd {\bf 49}, 3367 (1994).

\bibitem{voloshin97}
M. B.~Voloshin, \plb {\bf 397}, 275 (1997);
Z.~Ligeti, L.~Randall, and M.B.Wise,\plb {\bf 402}, 178 (1997);
A. K.~Grant, A.G.~Morgan, S.Nussinov, and R.D.~Peccei, \prd {\bf 56}, 3151 (1997);
G.~ Buchalla, G.~Isidori, and S. J.~Rey, \npb {\bf 511}, 594 (1998).

\end{thebibliography}
\end{document}